# High-pressure order-disorder transition in Mg$_2$SiO$_4$: Implications for super-Earth mineralogy

Rajkrishna Dutta[1,2*], Sally J. Tracy[1] and R. E. Cohen[1]

[1]Earth and Planets Laboratory, Carnegie Institution for Science, Washington DC 20015, USA.
[2]Discipline of Earth Sciences, Indian Institute of Technology Gandhinagar, Gujarat 382355, India.

**Abstract**

(Mg,Fe)SiO$_3$ post-perovskite is the highest-pressure silicate mineral phase in the Earth's interior. The extreme pressure and temperature conditions inside large extrasolar planets will likely lead to phase transitions beyond post-perovskite. In this work we have explored the high-pressure phase relations in Mg$_2$SiO$_4$ using computations based on density functional theory. We find that a partially disordered $I\bar{4}2d$-type structure would be stable under the conditions expected for the interiors of super-Earth planets. We have explored the mechanism of the phase transition from the ordered ground state and the effect of the disordering on electronic properties of the silicate phase. The discovery of a structure where two very dissimilar cations, Mg$^{2+}$ and Si$^{4+}$ occupy the same crystallographic site opens up a domain of interesting crystal chemistry and provides a foundation for other silicates and oxides with mixed occupancy.

**Introduction**

The lower mantle of the Earth is believed to be primarily composed of (Mg,Fe)SiO$_3$ perovskite (bridgmanite, Pv) [1]. With increasing pressure and temperature, Magnesium rich bridgmanite transforms to a layered CaIrO$_3$-type structure called post-perovskite (pPv) [2]. The properties of the pPv phase offer a probable explanation for several unusual properties (e.g. seismic anisotropy, topography and geochemical anomalies) of the D"-layer [3] at the base of the lower mantle. No further phase transitions are expected in the Earth. Recent advances in astronomy and planetary sciences have led to discovery of many exoplanets. Till date, more than

*Corresponding author: raj.dutta@iitgn.ac.in

5200 exoplanets [4] have been confirmed, among which > 1500 are super-Earth planets ($M_E < M < 10 M_E$). Super-Earth planets with a rocky interior are particularly significant because of their potential to harbor life. The mantles of these terrestrial exoplanets are likely made of earth forming silicate minerals with a range (0.67-1.5) of Mg/Si ratios [5,6]. The extreme pressure and temperature conditions (Fig. 1) inside these planets are expected to stabilize phases beyond post-perovskite. Knowledge of these phase transitions are crucial to understanding the dynamics and evolution of these planetary interiors.

The ultrahigh-pressure phase relations in the MgO-SiO$_2$ system have been extensively studied both experimentally and computationally. Laser-heated diamond cell experiments on Mg$_2$SiO$_4$ suggest MgSiO$_3$ pPv remains stable at least up to 265 GPa [7]. Higher pressure-temperature conditions remain inaccessible for even state of the art static compression techniques. Shock melting has been reported to occur at ~ 250 GPa for Mg$_2$SiO$_4$ [8], while MgSiO$_3$ glass melts at 180 GPa along the Hugoniot [9] and studies using bridgmanite starting material [10,6] observe shock melting at ~500 GPa. However, none of these experiments have reported a direct observation of pPv or any post-pPv phases.

Due to the challenges in accessing these extreme pressure and temperature conditions, our understanding of the mineralogy of terrestrial exoplanets are largely based on quantum mechanical computations. Computational studies [11–15] suggest pPv MgSiO$_3$ would ultimately dissociate into an assemblage of binary oxides after two or three stages of partial dissociation as follows:

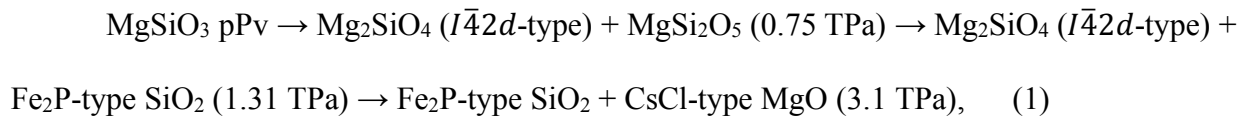

MgSiO$_3$ pPv → Mg$_2$SiO$_4$ ($I\bar{4}2d$-type) + MgSi$_2$O$_5$ (0.75 TPa) → Mg$_2$SiO$_4$ ($I\bar{4}2d$-type) + Fe$_2$P-type SiO$_2$ (1.31 TPa) → Fe$_2$P-type SiO$_2$ + CsCl-type MgO (3.1 TPa),    (1)

MgSiO$_3$ pPv + B1-MgO → Mg$_2$SiO$_4$ ($I\bar{4}2d$-type) (0.49 TPa) → Fe$_2$P-type SiO$_2$ + CsCl-type MgO (3.1 TPa),     (2)

MgSiO$_3$ pPv + pyrite-SiO$_2$ → MgSi$_2$O$_5$ (0.62 TPa) → Mg$_2$SiO$_4$ ($I\bar{4}2d$-type) + Fe$_2$P-type SiO$_2$ (1.25 TPa) → Fe$_2$P-type SiO$_2$ + CsCl-type MgO (3.1 TPa),     (3)

$I\bar{4}2d$-type MgSiO$_3$ was found computationally in compositions with a wide range of Mg/Si ratios [15]. Our recent experiments have explored the stability of this type of structure in NaMgF$_3$ [16] and Mg$_2$GeO$_4$ [17], well known analogs [18] of the MgO-SiO$_2$ system. The tetragonal $I\bar{4}2d$-type has not been experimentally observed in either case. In NaMgF$_3$, our experiments found the transition sequence:

NaMgF$_3$ (Pv) → NaMgF$_3$ (pPv) → NaMgF$_3$ (Sb$_2$S$_3$-type) → NaF (B2-type) + NaMg$_2$F$_5$ ($P2_1/c$) → NaF (B2) + MgF$_2$ (cotunnite-type),     (4)

Mg$_2$GeO$_4$ on the other hand was found to be stable in the thorium phosphide ($I\bar{4}3d$) structure or a disordered tetragonal $I\bar{4}2d$ phase, experimentally indistinguishable from the $I\bar{4}3d$ structure at pressures above ~175 GPa. In this study, we have explored the stability and nature of the $I\bar{4}2d$- and $I\bar{4}3d$-type phases in Mg$_2$SiO$_4$ using density functional theory (DFT) computations.

**Computational details**

We have performed plane wave density functional theory computations as implemented in the QUANTUM ESPRESSO [19] package. All computations were performed with the PBE exchange correlation functional and GBRV potentials [20]. An energy cutoff of 40 Ry was used for the orbitals. The Brillouin zone for the pPv and the ordered $I\bar{4}2d$-type structures were sampled with a 6x6x6 $k$-point grid.

To understand the behavior of the order-disorder transition, we considered an order parameter, Q. The order parameter was varied from 0 (completely disordered) to 1 (completely ordered) with thirteen intermediate values using the following relation: $X_{Mg}(8d) = Q \times 1/3 + 2/3$ and $X_{Mg}(4a) = -Q \times 2/3 + 2/3$. The disordered structures were generated using the ATAT toolkit [21] following the Special Quasirandom Structure method [22,23]. For each Q, we generated a 224 atom (2x2x2) supercell with the lattice parameters of the cubic disordered phase doubled in each direction. For each supercell, the objective function which attempts to match the maximum number of correlation functions of the SQS structure and the disordered state is calculated considering clusters of 2 cation pairs up to 5.0 Å. The objective function is then minimized using a Monte Carlo method. The best (lowest objective function) SQS structure was then optimized at the required pressure by relaxing both lattice parameters and atomic positions using the Broyden-Fletcher-Goldfarb-Shanno algorithm till the forces were $< 5 \times 10^{-4}$ Ry/Bohr. All relaxations were carried out at the $\Gamma$ point with a 40 Ry energy cutoff. To test the convergence of the calculations, we also performed a couple of optimizations with different *k*-point grids and energy cutoffs (Table S1 of the supplementary material). The same exercise was repeated for 13 configurations of order parameter from $Q = 0$ ($I\bar{4}3d$) to $Q = 1$ ($I\bar{4}2d$). The variation in enthalpy with Q could be fit well with a second-degree polynomial, no higher order fits were necessary. The disordering enthalpy (H) is counteracted by the configurational entropy ($\frac{S}{k_B} = \sum X \ln X$), where the sum is over the sites, X is the mole fraction of Mg or Si on each site and $k_B$ is the Boltzmann constant. The free energy ($H - TS$) is minimized at each temperature to obtain the variation in Q with temperature. A Mathematica [24] script, sample input file and instructions for this calculation has been provided with the supplementary material. For the density of state (DOS) computations, a finer 4x4x4 *k*-point grid was used.

**Results**

Initially, we optimized the structures of $I\bar{4}2d$-type Mg$_2$SiO$_4$, pPv MgSiO$_3$ and MgO (B1, B2) at select pressures. The structure of post-perovskite MgSiO$_3$ (space group: *CmCm*) consists of edge and corner sharing silicate octahedra (coordination number, CN = 6) forming layers along the *b*-axis. B1 ($Fm\bar{3}m$) and B2 ($Pm\bar{3}m$) MgO have the six coordinated rocksalt and CsCl-type structures respectively. In agreement with existing experiments [25] and *ab initio* calculations [26], we find B1-MgO transforms into B2-MgO at 525 GPa. Post-perovskite MgSiO$_3$ + MgO (B1/B2) is found to recombine into $I\bar{4}2d$-type Mg$_2$SiO$_4$ at 535/545 GPa (Fig. 2). Considering the different functional (PBE) used in this study, this is in reasonable agreement with Umemoto et al., 2017 (490 GPa using local density approximation). The reaction line to form $I\bar{4}2d$-type Mg$_2$SiO$_4$ has a negative Clapeyron slope (-16 MPa/K) at the pressure and temperature conditions [15] expected in super-Earth mantles. $I\bar{4}2d$-type Mg$_2$SiO$_4$ has a body-centered tetragonal structure with two different cation sites: Mg (*8d*) and Si (*4a*). Both Mg and Si show eight-fold coordination. The ordered $I\bar{4}2d$-type structure was not observed experimentally in Mg$_2$GeO$_4$ [17], a widely used analog of Mg$_2$SiO$_4$. Instead, an intrinsically disordered $I\bar{4}3d$-type or a highly disordered $I\bar{4}2d$-type structure was found to be stable at pressures > 175 GPa. Recent studies on Mg$_2$GeO$_4$ [17, 27] show that the $I\bar{4}3d$ phase can be understood in terms of an order-disorder transition from the $I\bar{4}2d$ structure. In the $I\bar{4}3d$-type structure, the Mg and Si coordination remains unchanged, but the cations occupy a single site (*12a;* Mg: 2/3, Ge: 1/3).

Figure 3 shows the structure (224 atom supercell) of partially disordered (Q = 0.53125) and completely disordered Mg$_2$SiO$_4$. The nature and order of the transition [28] is dependent on the behavior of enthalpy, *H(Q)* in comparison to entropy, *S(Q)*. The order parameter can abruptly change with temperature (first order) or show a continuous change with no phase transformation.

Figures 4a and 4b show the change in enthalpy and entropy as a function of the order parameter respectively at 600, 800 and 1500 GPa. The enthalpy of disorder increases by 0.012 Ry/atom, 0.014 Ry/atom and 0.018 Ry/atom at 600, 800 and 1500 GPa respectively on complete disordering (Q = 1→ Q = 0). The configurational entropy (p.f.u) is zero in the completely ordered $I\bar{4}2d$ structure and 1.9095 in the completely disordered $I\bar{4}3d$ structure.

Figure 5 shows the change in the order parameter with temperature. Q is found to vary smoothly with temperature, suggesting a gradual disordering with temperature, without a phase transition. At 600 GPa, disordering is found to start at 1100K (Q = 0.999), with continuous decrease in Q to the peak temperature of 20,000 K (Q = 0.056). At higher pressures, the amount of disorder is less at a particular temperature, suggesting a positive Clapeyron slope between the ordered and disordered structures. This is in agreement with the transition slope reported in $Mg_2GeO_4$ [27]. In this case, the transition is isostructural within the $I\bar{4}2d$ space group, from a highly ordered to a moderately disordered structure. The ordering is convergent because the two cation sites in the tetragonal structure eventually becomes equivalent leading to a change in symmetry (cubic). The disordering process becomes sluggish at high Q values as can be seen from the nearly horizontal slope at Q < 0.15. This can be explained using the relative changes in enthalpy of disorder and configurational entropy. The ΔS ordering is fairly flat (Fig. 4b) from Q = 0.2 to Q = 0, whereas $\Delta H$ continues to increase (Fig. 4b), so it does not pay to order completely, even at extremely high temperatures (>20,000 K).

Figures 6a and 6b show the change in unit-cell volume and the *c/a* ratio as a function of the order parameter. Disordering does not have a significant effect on the cell volume, showing a 0.6%, 0.5% and 0.3% decrease over the entire Q range at 600, 800 and 1500 GPa respectively.

The *c/a* ratio changes from 1 in the cubic phase to 1.036, 1.030 and 1.022 respectively in the completely ordered tetragonal structure.

At pressure and temperature conditions relevant to the interior of giant super-Earth planets, the mineral phases are expected to be semi-conductors with band gaps [11]. Fig. S1 shows the electronic band gap of $Mg_2SiO_4$ as a function of pressure and order parameter. Fig. S2 shows the density of states (DOS) of the completely ordered and completely disordered $I\bar{4}2d$-type structure at 800 GPa. Pressure causes both the conduction and valence band to shift to higher energies, with the conduction band increasing at a faster pace in comparison to the valence band. This leads to a slight increase in the band gap with pressure at a given Q (Fig. S3). Disordering on the other hand, has a significant effect on the electronic structure of $Mg_2SiO_4$. The peak values of the DOS in both the valence and conduction bands decrease from Q = 1 to Q = 0, suggesting an increase in the overlap of the bonds. The calculated band gaps (which probably grossly underestimate the true gaps) of the completely ordered $I\bar{4}2d$-type phase is 10.44, 10.74 and 10.89 eV at 600, 800 and 1500 GPa respectively. This is significantly higher than the reported values for $MgSiO_3$ and the binary oxides at the respective pressures. A complete disorder of the silicate would lead to a ~61% reduction in the band gap at considered pressures.

**Discussion and Conclusion**

In this study, we have shown that $MgSiO_3$ post-perovskite + MgO will recombine into $I\bar{4}2d$-type $Mg_2SiO_4$ at pressures > 535 GPa. Our computations suggest that the ordered tetragonal $I\bar{4}2d$-type structure will gradually disorder with increasing temperature. For a terrestrial super-Earth with mass equal to 10 $M_E$, the pressure and temperature at the core-mantle boundary is likely ~1300 GPa and 6500 K [29], while the melting point of $MgSiO_3$ is expected to be ~13000 K at 1500 GPa. Thus, a complete disorder into the cubic $I\bar{4}3d$-type structure is not expected at conditions relevant to planetary interiors, instead a partially disordered $I\bar{4}2d$-type $Mg_2SiO_4$ is potentially the stable silicate phase.

In a recent work [17], we explored the high-pressure behavior of $Mg_2GeO_4$, a very good analogue for $Mg_2SiO_4$. The germanate mirrors the phase transition sequence [2,30] in the silicate (Pv → pPv at 125 GPa, pPv + MgO → $Mg_2GeO_4$ at 535 GPa), but at substantially lower pressures (Pv → pPv at 65 GPa, pPv + MgO → $Mg_2GeO_4$ at ~185 GPa). Figure 5 compares the change in order parameter with temperature in $Mg_2SiO_4$ and $Mg_2GeO_4$. The computationally predicted ground state structure i.e. the ordered $I\bar{4}2d$-type structure was not experimentally observed in the germanate. Instead, a disordered $I\bar{4}3d$-type or a highly disordered (Q > 0.53) $I\bar{4}2d$-type structure, experimentally indistinguishable (i.e. the peak splittings expected for the tetragonal structure cannot be experimentally resolved) from the completely disordered $Th_3P_4$-type structure was found to be consistent with the X-ray diffraction patterns. $Mg_2SiO_4$ on the other hand is expected to disorder less at pressures close to the pPv + MgO → $I\bar{4}2d$ transition.

The perovskite to post-perovskite phase transition in $MgSiO_3$ is accompanied by a 1-1.5% reduction in volume [31]. The post-post-perovskite transition to the partially disordered $I\bar{4}2d$-type structure is computed to have a 0.8 – 1.4% change at 600 GPa. The large negative Clapeyron slope [15] of the transition combined with a modest volume change can lead to a possible boundary layer in the super-Earth mantles.

The $Th_3P_4$-type or modified $I\bar{4}2d$ structure is common in $A_2X_3$, $A_3X_4$ and $AB_2X_4$ rare earth chalcogenides [32,33]. The structure has found widespread technical applications because its highly flexible and can incorporate defects and impurities [34]. A recent study also reported a modified $Th_3P_4$-type structure in $Fe_3O_4$ at 78 GPa and 4800 K [35,36]. The structural flexibility and widespread occurrence of this structure in chalcogenides, oxides and silicates/ germanates makes it a possible important transition pathway. The intrinsically disordered nature of this phase has important implications chemical miscibility at high P-T conditions and may lead to anomalous thermal conductivities.

**Acknowledgements**

RD gratefully acknowledges support from the Carnegie Endowment and IIT Gandhinagar. Computations were performed using the GCS Supercomputer SuperMUC-NG at Leibniz Supercomputing Centre (LRZ, www.lrz.de).

Figure 1. Theoretically computed phase boundaries (solid black lines) in the Mg$_2$SiO$_4$ system [15]. Dashed black lines show the pressure and temperature conditions inside terrestrial exoplanets [29] with masses equal to 5 and 10 times that of the Earth ($M_E$). Solid red line shows the melting curve of MgSiO$_3$ [6]. Red rectangles indicate estimated core-envelope P, T conditions in giant planets in the solar system.

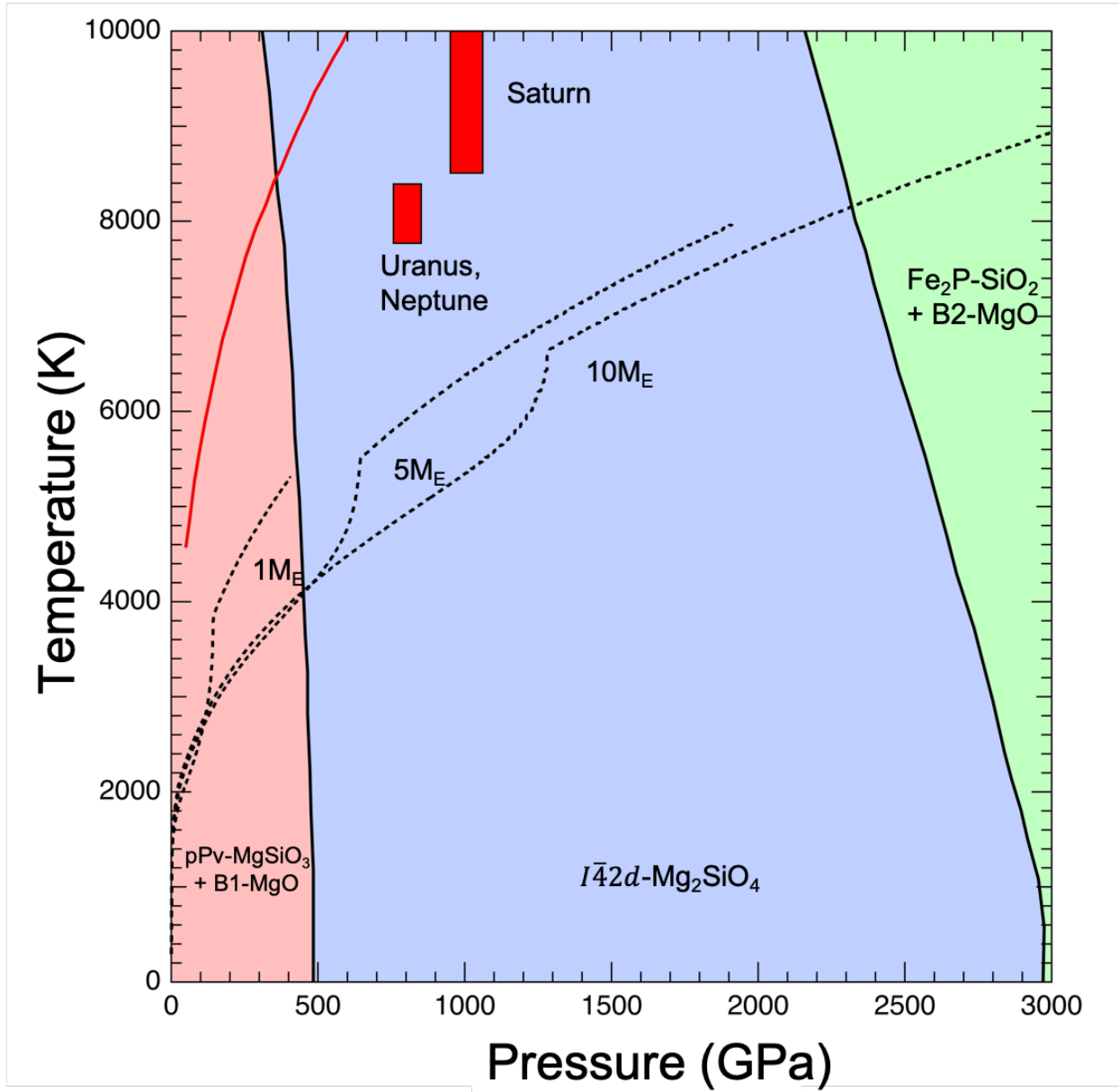

Figure 2. Enthalpy differences ($\Delta H$) between post-perovskite MgSiO$_3$ + MgO (B1/B2) and $I\bar{4}2d$-type Mg$_2$SiO$_4$ as a function of pressure for the static lattice (classical 0 K). The dashed black line indicates $\Delta H = 0$; the crossover with the respective phases shows the transition pressure.

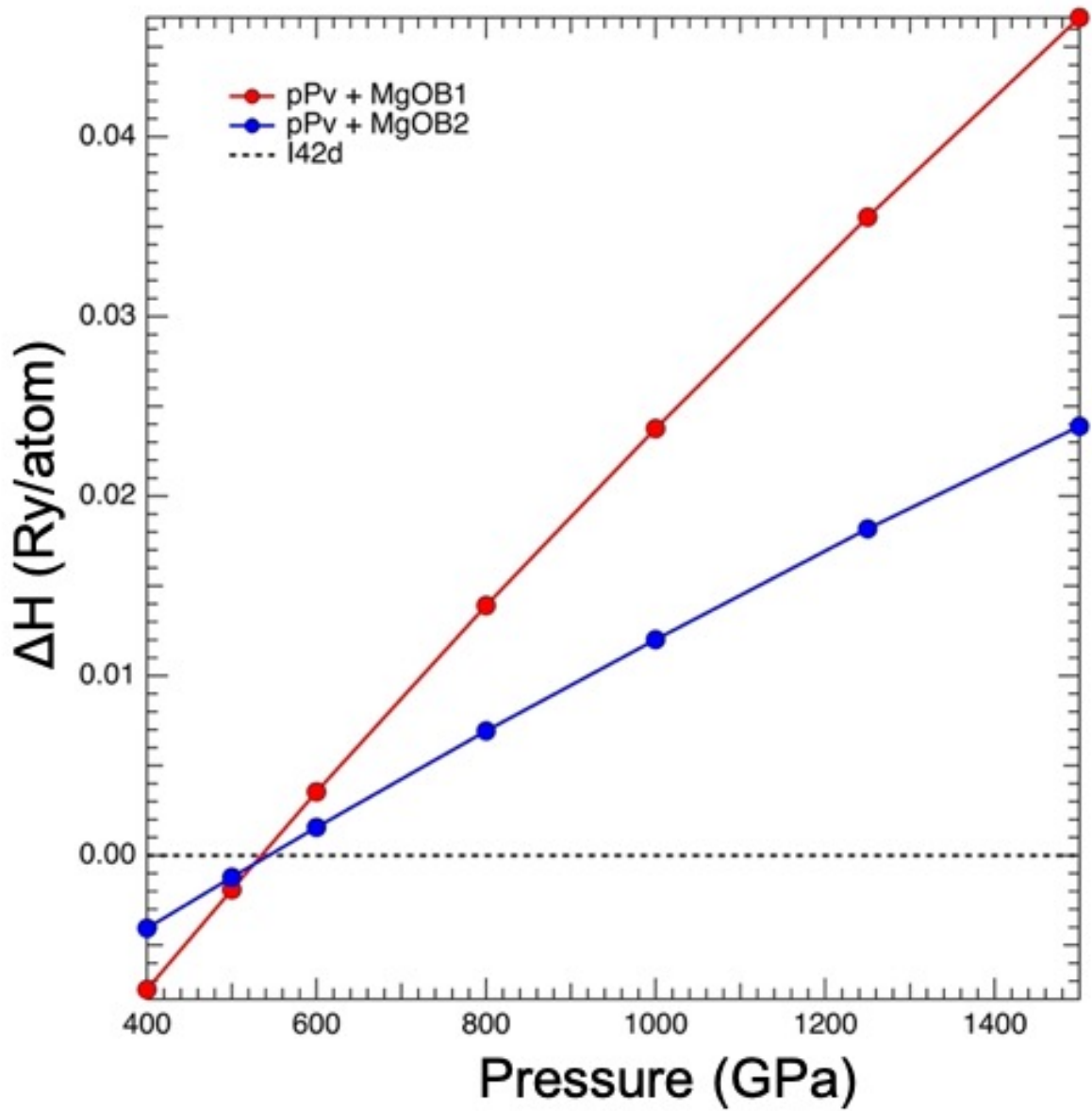

Figure 3. Crystal structure (224 atoms supercell) of partially and completely ordered $I\bar{4}2d$-type Mg$_2$SiO$_4$ at 600 GPa.

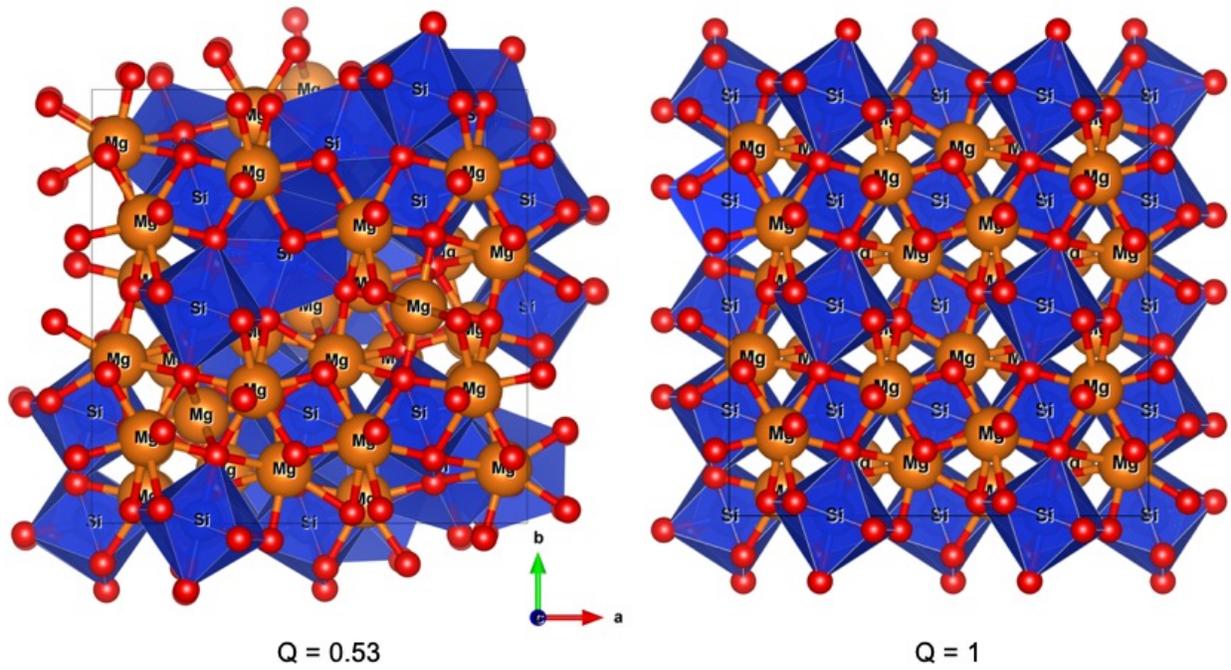

Figure 4. Difference in enthalpy ($H_Q - H_{Q1}$) (left) versus order parameter, $Q$ at 600, 800 and 1500 GPa; where $H_{Q1}$ is the enthalpy of the completely ordered phase ($Q = 1$). Configurational entropy at 600 GPa (right).

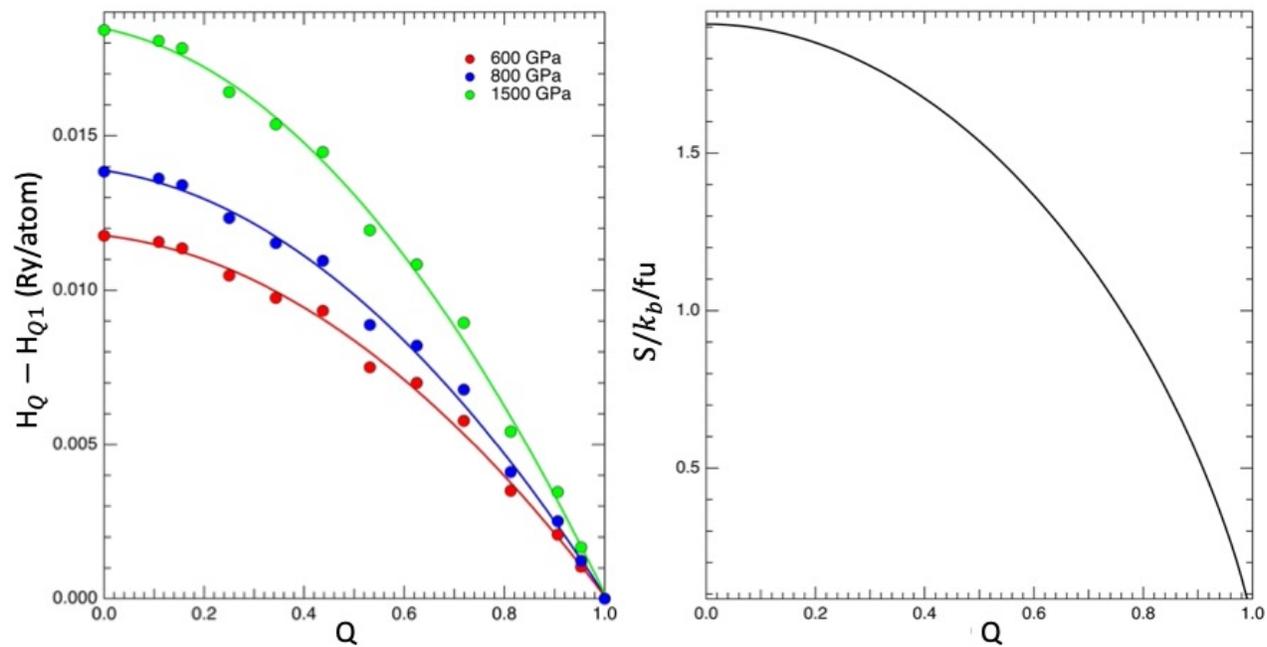

Figure 5. Change in the order parameter, $Q$ with temperature for $I\bar{4}2d$-type $Mg_2SiO_4$ (solid) and $Mg_2GeO_4$ (dashed) [17] at select pressures. $Q = 0$ and $Q = 1$ corresponds to completely disordered and completely ordered structures respectively. Black dotted line shows the predicted melting temperature [6] of $MgSiO_3$ based on extrapolation of shockwave meting data.

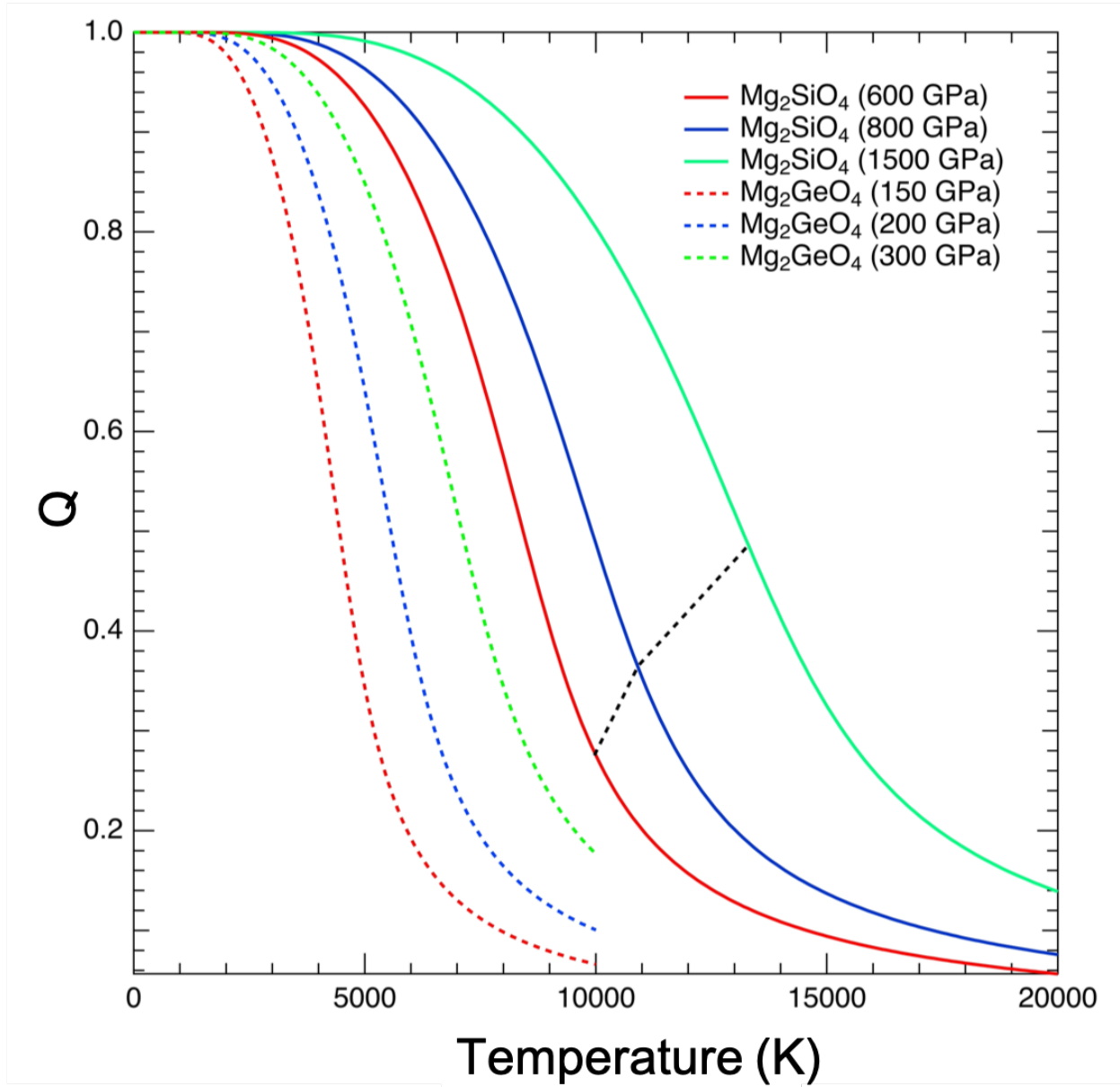

Figure 6. Variation in unit cell volume (left) and $c/a$ of $I\bar{4}2d$-type $Mg_2SiO_4$ as a function of the order parameter $Q$ at 600, 800 and 1500 GPa.

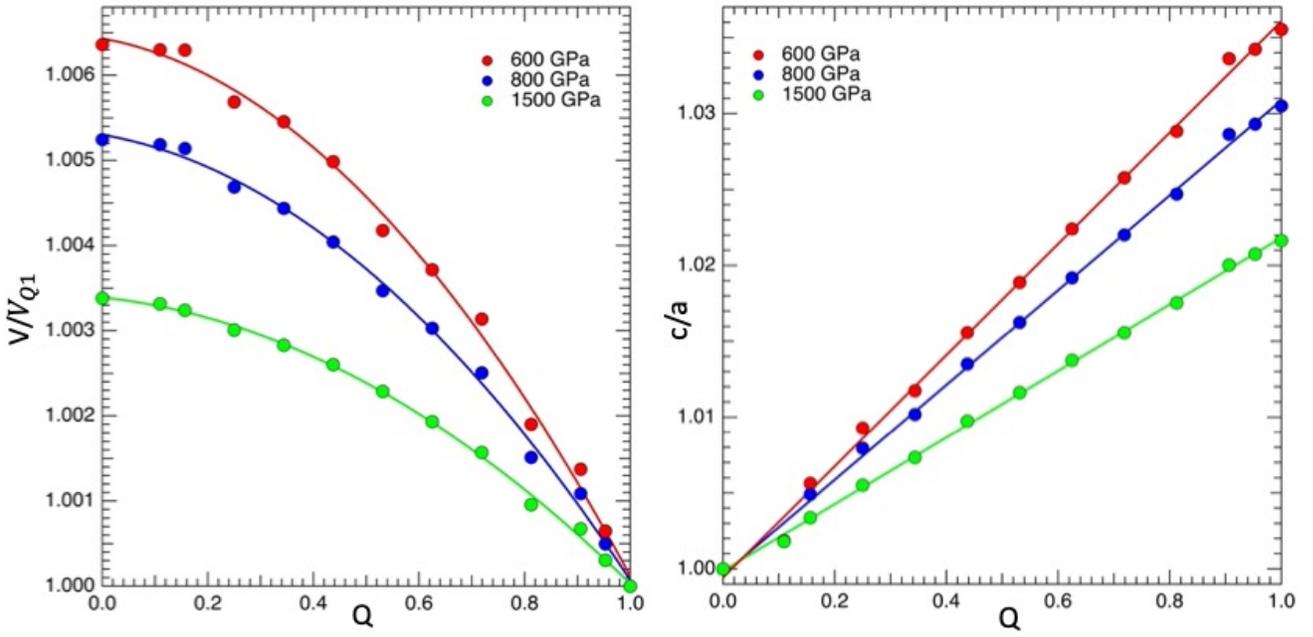

# High-pressure order-disorder transition in $Mg_2SiO_4$: Implications for super-Earth mineralogy

Rajkrishna Dutta[1,2*], Sally J. Tracy[1] and R. E. Cohen[1]

[1]Earth and Planets Laboratory, Carnegie Institution for Science, Washington DC 20015, USA.
[2]Discipline of Earth Sciences, Indian Institute of Technology Gandhinagar, Gujarat 382355, India.

Figure S1. Computed band gap of $I\bar{4}2d$-type $Mg_2SiO_4$ at different order parameters as a function of pressure (red – 600 GPa, blue – 800 GPa and green – 1500 GPa). The black lines show the band gaps of pPv-$MgSiO_3$, B2-MgO and cotunnite-type $SiO_2$ [11].

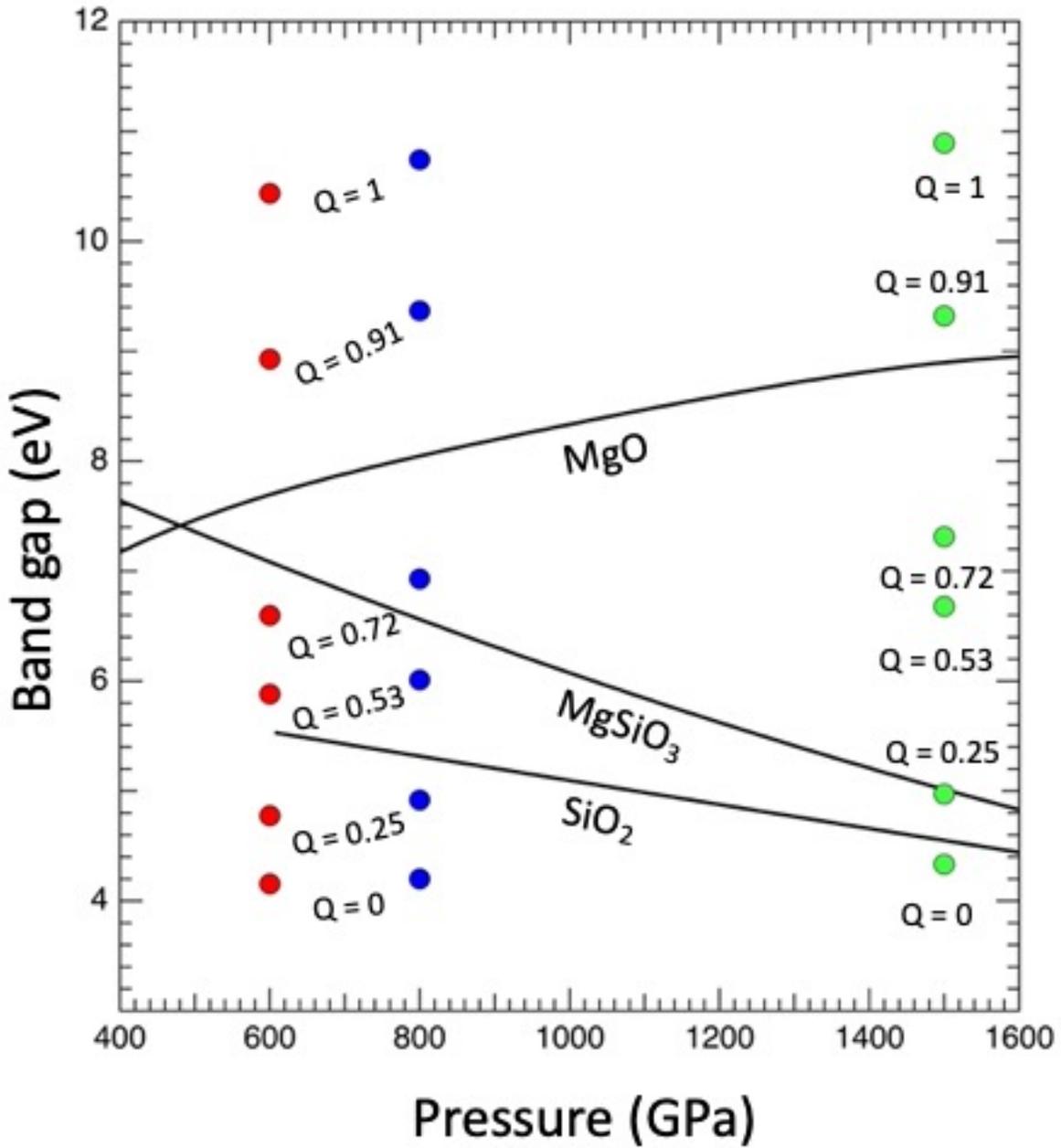

Figure S2. Total density of states of completely ordered and completely disordered $I\bar{4}2d$-type Mg$_2$SiO$_4$ at 800 GPa. E$_F$ denotes the Fermi Level.

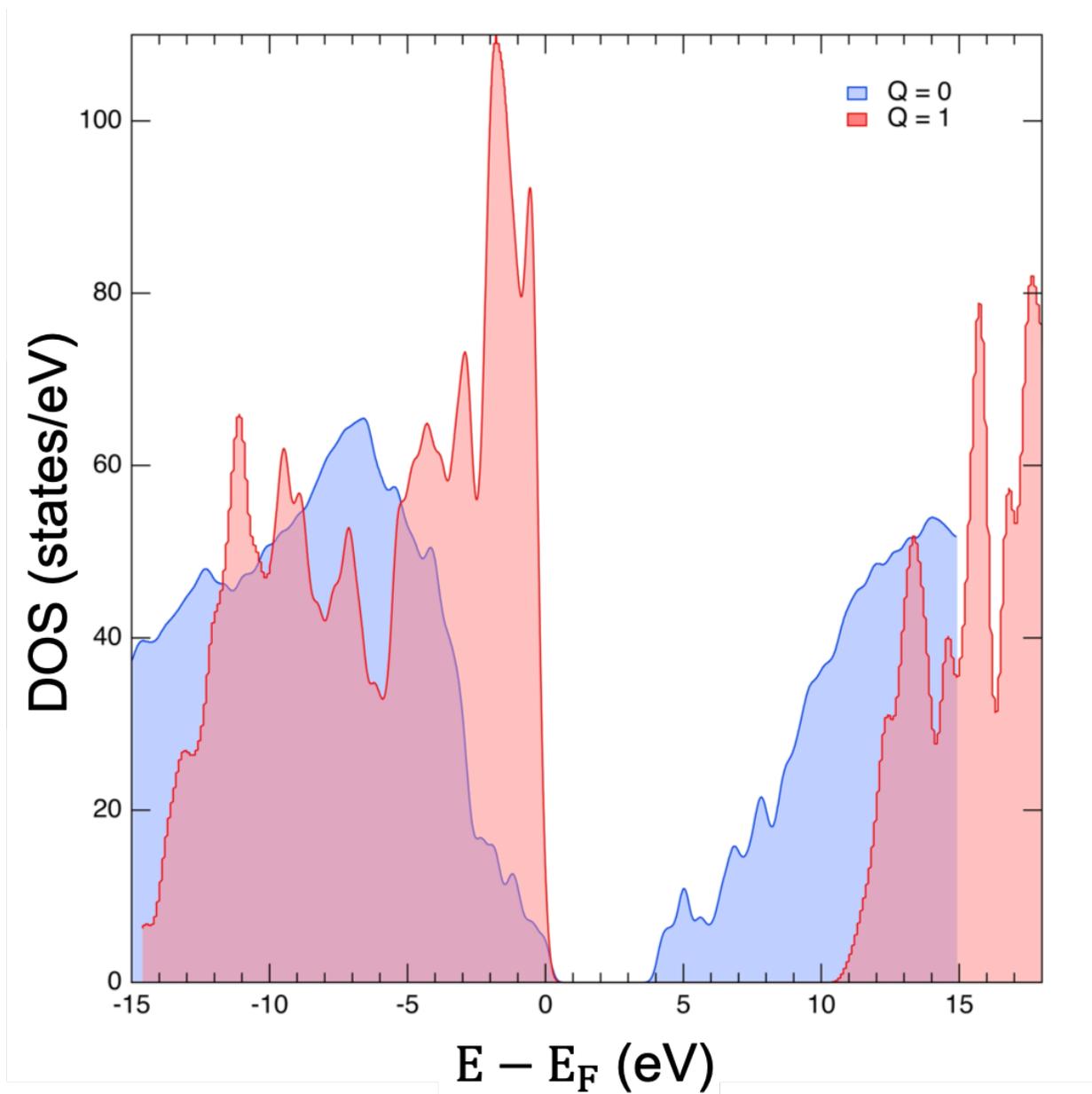

Figure S3. Density of states of $I\bar{4}3d$-type $Mg_2SiO_4$ (Q = 0) at select pressures.

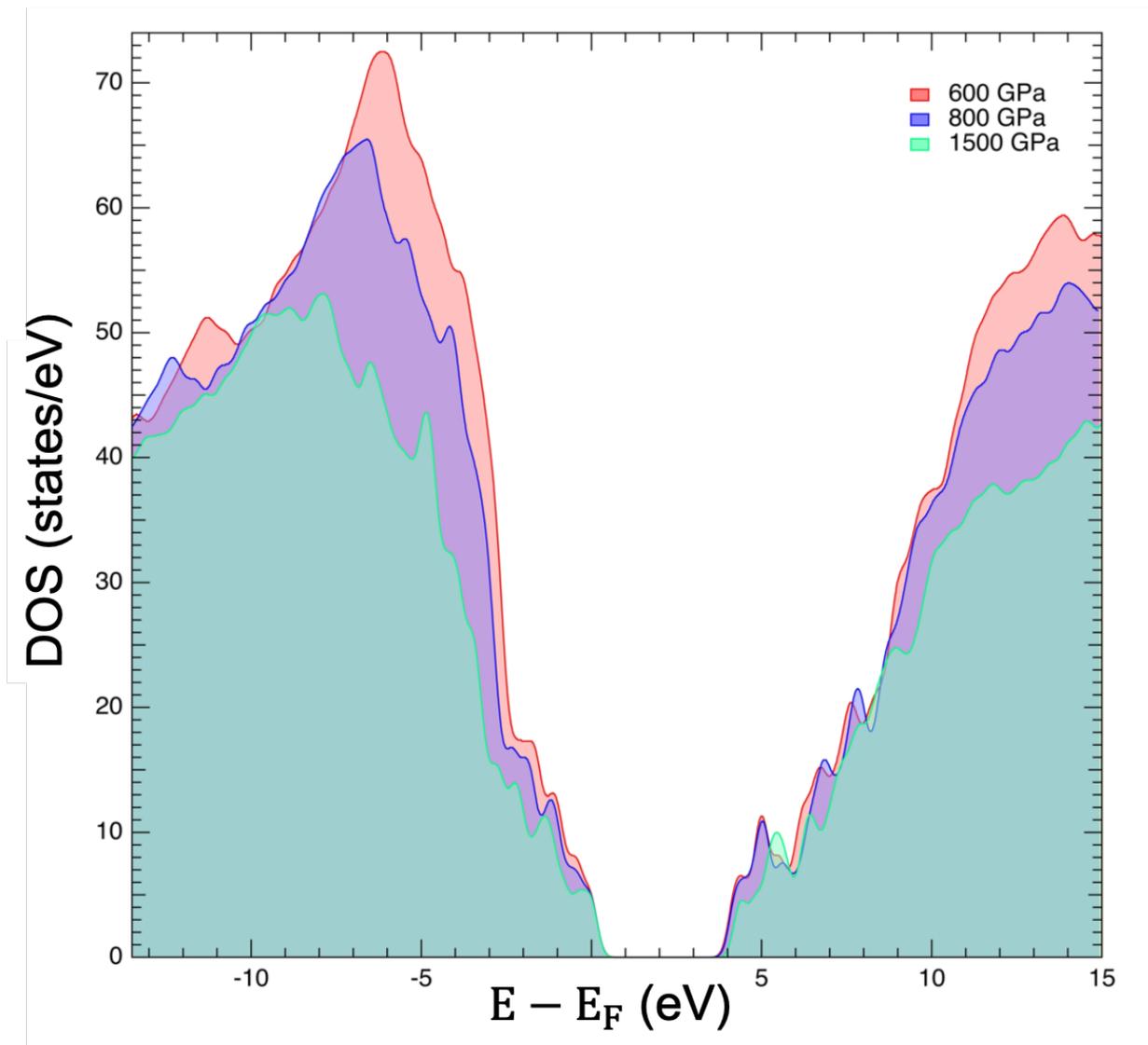

Table S1. Enthalpy difference, $\Delta H = (H_{Q_{0.25}} - H_{Q_{0.625}})$ at 600 GPa with different cutoff energies and *k*-point grids.

| Energy cutoff (Ry) | k-point | ΔH (Ry/atom) |
|---|---|---|
| 40 | Γ | 0.003477151 |
| 100 | Γ | 0.003476579 |
| 40 | 1x1x1 | 0.003468776 |
| 40 | 2x2x2 | 0.00347197 |
| 100 | 2x2x2 | 0.003472757 |
| 40 | 4x4x4 | 0.003471417 |